\documentstyle[12pt]{article}

\newcommand{\be}{\begin{equation}}
\newcommand{\ee}{\end{equation}}
\newcommand{\bel}[1]{\begin{equation}\label{#1}}
\newcommand{\bea}{\begin{eqnarray}}
\newcommand{\eea}{\end{eqnarray}}
\newcommand{\ba}{\begin{array}}
\newcommand{\udl}{\underline}

\newcommand{\ket}[1]{\mbox{$| \, {#1}\, \rangle$}}
\newcommand{\exval}[1]{\mbox{$\langle \, {#1}\, \rangle$}}

\begin{document}

\begin{titlepage}
\thispagestyle{empty}
\begin{center}
{\Large \bf
Dynamic Matrix Ansatz for Integrable Reaction-Diffusion Processes
}\\[13mm]

{\large {\sc
Gunter M. Sch\"utz}
} \\[8mm]

\begin{minipage}[t]{13cm}
\begin{center}
{\small\sl
Institut f\"ur Festk\"orperforschung,\\
Forschungszentrum J\"ulich, 52425 J\"ulich, Germany\\
e-mail: g.schuetz@kfa-juelich.de
}
\end{center}
\end{minipage}
\vspace{15mm}
\end{center}
{\small We show that the stochastic dynamics of a large class of
one-dimensional interacting particle systems may be presented by integrable
quantum spin Hamiltonians. Generalizing earlier work \cite{Stin95a,Stin95b}
we present an alternative description of these processes in terms of a
time-dependent operator algebra with quadratic relations. These relations
generate the Bethe ansatz equations for the spectrum and turn the
calculation of time-dependent expectation values into the
problem of either finding representations of this algebra or of solving
functional equations for the initial values of the operators. We use
both strategies for the study of two specific models:
(i) We construct a two-dimensional time-dependent representation of the
algebra for the symmetric exclusion process with open boundary conditions.
In this way we obtain new results on the dynamics of this system and on
the eigenvectors and eigenvalues of the corresponding quantum spin
chain, which is the isotropic Heisenberg ferromagnet with non-diagonal,
symmetry-breaking boundary fields. (ii) We consider the non-equilibrium
spin relaxation of Ising spins with zero-temperature Glauber dynamics and
an additional coupling to an infinite-temperature heat bath with Kawasaki
dynamics. We solve the functional equations arising from the algebraic
description and show non-perturbatively on the level of all finite-order
correlation functions that the coupling to the infinite-temperature heat
bath does not change the late-time behaviour of the zero-temperature
process. The associated quantum chain is a non-hermitian anisotropic
Heisenberg chain related to the seven-vertex model.
}
\vspace{6mm}\\
\udl{PACS numbers:} 05.40+j, 02.50Ga, 05.70.Ln, 75.40.Gb
\end{titlepage}
\newpage

\section{Stochastic Dynamics and Quantum Systems}

One-dimensional stochastic reaction-diffusion processes are of both theoretical
and experimental interest in a very wide context. They are well-known
models both for reaction-diffusion mechanisms in physics and
chemistry and for stochastic spin flip dynamics \cite{Priv97}.
More recently they have appeared through various mappings also as
models for traffic flow \cite{Scha93,Yuka94,Schr95,Naga95,Raje97}, the kinetics
of biopolymerization \cite{MacD68,Schu97a}, reptation of DNA in gels
\cite{Wido91,vanL92,Bark94,Bark96a,Bark96b,Prae96}, interface growth
\cite{Krug91b,Halp95}, diffusion in zeolites
\cite{Kukl96,Roed97} and many other phenomena. Even in relatively simple models
of driven diffusion such as the asymmetric exclusion process with open
boundaries one finds
a very rich dynamical behaviour involving dynamical and non-equilibrium phase
transitions of various kinds \cite{Krug91a,Schu93b,Schm95,Kolo97}.
Exact solutions \cite{Priv97,Schu98} allow for a detailed understanding
of cooperative phenomena in these classical many-body systems and provide
insight in the role of inefficient diffusive mixing in diffusion-limited
chemical reactions, in the dynamics of shocks and in other fundamental
mechanisms which determine the behaviour of low-dimensional systems
far from thermal equilibrium.

A convenient and much used description of stochastic processes is in terms
of a master equation for the probability distribution $f(\udl{n};t)$ of
the stochastic variables $\udl{n}$. These variables represent the states
in which the system may be found at any given instant of time. The master
equation encodes the transition probabilities $p(\udl{n}' \to \udl{n})$ of
moving from one state $\udl{n}'$ to another state $\udl{n}$ in one
time step: the probability distribution $f(\udl{n};t+\Delta t) =
\sum_{\udl{n}'} p(\udl{n}' \to \udl{n}) f(\udl{n'};t)$ is just
the sum of probabilities of finding the system in state $\udl{n'}$
at time $t$ times the respective transition probabilities
$p(\udl{n}' \to \udl{n})$. Thus the
master equation expresses the probability of finding the system at time
$t+\Delta t$ in a given configuration $\udl{n}$ in terms of the probability
distribution at time $t$. Such processes are Markov
processes which may be constructed for the description of interacting
particle systems \cite{Ligg85}.

Since the master equation is linear in the probability
distribution, it can be expressed as a vector equation in a ``quantum
Hamiltonian formalism'' by mapping each state $\udl{n}$ of the system to a
basis
vector $\ket{\udl{n}}$ in a suitable vector space $X$. Thus the
probability distribution becomes a vector $|\,f(t)\,\rangle
=\sum_{\udl{n}} f(\udl{n};t) \ket{\udl{n}}$ and the
master equation for a continuous-time process ($\Delta t \to 0$)
takes the form of a Schr\"odinger equation in imaginary time
\be
\frac{d}{d t} |\,f(t)\,\rangle = - H |\,f(t)\,\rangle.
\label{1}
\ee
For various interesting many-body systems the generator
$H$ of the (classical) stochastic time evolution turns out to be identical to
the quantum Hamiltonian of spin chains known from condensed matter
physics \cite{Sigg77,Alex78,Gwa92,Sand93,Alca93,Schu95} (see next section).
The ground state of a stochastic Hamiltonian (which by construction has energy
0) corresponds to the stationary probability distribution of the stochastic
dynamics which is reached asymptotically at the late stages of the time
evolution. Through this somewhat surprising connection to quantum spin systems
the stochastic dynamics become tractable with the tool box of quantum
mechanics and progress may be achieved.\footnote{For
discrete time dynamics the vector form of the master equation reads
$|\,f(t+\Delta t)\,\rangle = T |\,f(t)\,\rangle$. For many interesting
systems in one dimension $T$ is the transfer matrix of a two-dimensional
vertex model \protect\cite{Kand90,Schu93a,Raje96,Hone97}. The
frequently-used notion
``quantum Hamiltonian formalism'' is somewhat misleading in so far
as for many stochastic applications the transition rates result in coupling
constants which make $H$ non-hermitian. Further, since $f$ is a
(real-valued) probability and not a (complex) probability amplitude, the
expectation values of the stochastic process are not the expectation
values normally calculated in a quantum mechanical problem.}

A paradigmatic example of this relationship is the symmetric exclusion process
\cite{Ligg85,Spit70}. In this lattice model particles hop between
lattice sites $k,l$ with rates $p_{k,l}=p_{l,k}$. They interact via a
hard-core repulsion which prevents the occupation of a lattice site by
more than one particle. The stochastic quantum Hamiltonian obtained for
this system \cite{Alex78} is the
Hamiltonian for the isotropic spin-(1/2) Heisenberg ferromagnet
\bel{Heis}
H = - \frac{1}{2} \sum_{k,l} p_{k,l}  \left[ \sigma^x_k\sigma^x_l
               + \sigma^y_k\sigma^y_l + \sigma^z_k\sigma^z_l -1 \right].
\ee
The manifest $SU(2)$-symmetry of $H$ (which is not recognizable in the original
master equation) can be used for obtaining exact results \cite{Gwa92,Schu94}.
Moreover, in one dimension, the system with
nearest neighbour hopping is integrable and can be solved by the Bethe
ansatz \cite{Beth31}.

The integrability is not a special feature of the symmetric exclusion process
alone. Driven diffusion in one dimension
($p_{k,k+1}/p_{k+1,k}=q \neq 1$) is
described by the $XXZ$ quantum spin chain which
differs from (\ref{Heis}) by an anisotropy $\Delta = (q+q^{-1})/2$
in the $z-z$ coupling of the spins \cite{Gwa92}. Using the Bethe ansatz
and related methods many new exact results particularly on the dynamical
properties of the process have been obtained
\cite{Gwa92,Sand94a,Henk94,Kim95,Schu97b,Derr98,Alim98}. Moreover, it turns out
that a 10-parameter class of reaction-diffusion systems of identical
particles \cite{Schu95} and  various systems of non-identical
particles \cite{Alca93,Dahm95} are described by integrable quantum chains.
Unfortunately however, the Bethe ansatz requires knowledge of some reference
eigenstate of $H$ and integrability has so far failed to provide constructive
methods to calculate even the ground state of spin chains with
boundary fields that destroy the reference state. Such boundary fields
are of importance in stochastic dynamics of non-equilibrium systems as they
allow for a modelling of open systems which are
in contact with external particle reservoirs at their boundaries and thus allow
for maintaining a particle current through the system.

In a very different approach the ground states of one-dimensional spin
Hamiltonians are formulated in terms of {\em matrix product states}
\cite{Affl88,Fann88,Klum91,Klum93} where the ground state wave function
is expressed in terms of a trace over a product of matrices. They may be
seen as representations of an operator algebra which is determined by
the requirement that by acting with the Hamiltonian on this state one obtains
an eigenstate
(the ground state) of $H$. Applied to stochastic Hamiltonians one obtains
in this way the stationary distribution of a stochastic process
\cite{Derr93,Sand94b,Essl96,Hinr96a,Kreb97,Arnd97}.\footnote{For non-periodic
systems with boundary fields one does not take a trace, but a suitably
chosen scalar product \protect\cite{Derr93}.}

By constructing an infinite-dimensional representation of the stationary
matrix algebra for the asymmetric exclusion process with open boundaries
Derrida et al. \cite{Derr93} produced the same exact results that were
obtained independently by Sch\"utz and Domany
\cite{Schu93b} using a different method. In fact, with hindsight
our treatment may be seen as a representation-free solution of recursion
relations that one can derive from the matrix algebra. It is the
aim of this work to apply both strategies, viz. (i) construction of a matrix
representation or (ii) solution of equations resulting from the algebraic
relations alone, to a matrix product treatment of the {\em dynamics} of
reaction-diffusion systems. Specifically we consider
the symmetric exclusion process with open boundaries and a
reaction-diffusion mechanism which is equivalent to a spin relaxation model.

The extension of the stationary matrix approach of Derrida et al. to a
dynamical description requires one new idea. This is the introduction
of auxiliary matrices $S,T$ \cite{Stin95a,Stin95b} which do not appear in the
calculation of expectation values, but are necessary to formulate the
dynamical algebra which is determined such that the probability distribution
satisfies the full time-dependent master equation.
The special case of this construction where $S+T=0$ allowed
among other things for the rederivation \cite{Stin95b} of the Bethe ansatz
equations \cite{Alca87} for the spectrum in the symmetric case with open
boundaries and for the rederivation \cite{Sasa97a} of the spectrum
in the asymmetric case with periodic boundary conditions \cite{Kim95}.
With the auxiliary matrices as additional ingredient the extension of the
dynamical matrix product approach to reaction-diffusion systems becomes
straightforward \cite{Schu96} (see below).

I will first show (Sec. 2) how reaction-diffusion systems of identical
hard-core particles are related to a generalized Heisenberg chain.
Its spectrum can be obtained from the Bethe ansatz. This is a simplified
rederivation of some results obtained earlier \cite{Schu95}. Then in
(Sec.~3) I will generalize the operator approach to the general
reaction-diffusion problem of identical hard-core particles with nearest
neighbour interaction in one dimension. As an application I will return to
the symmetric exclusion process and present a two-dimensional
representation of the time-dependent operator algebra.\footnote{Sections 2
and 3 of this paper are not really new. They constitute the bulk of the
paper \cite{Schu96} which was presented at the Satellite Meeting to
Statphys 19 on Statistical Models, Yang-Baxter Equation and Related Topics,
at Nankai University, Tianjin (August 1995).} In Sec.~4 I will solve a
non-equilibrium spin relaxation model introduced by Droz et al.
\cite{Droz89} by solving functional equations arising from the algebraic
relations of the time-dependent matrix algebra. This treatment does not
require the calculation of representations of the algebra. In Sec.~5 the
main results are summarized and some open questions are pointed out.

\section{Integrable Reaction-Diffusion Processes}

We will consider stochastic reaction-diffusion processes of identical particles
with hard-core repulsion moving on a ring with $L$ sites. Even though part of
our approach generalizes to arbitrary lattices \cite{Schu95} we will study here
only one-dimensional systems with nearest neighbour interaction. The stochastic
variables of the system are the occupation numbers $\underline{n} = \{ n_k \}$
where $n_k=0,1$ indicates whether site $1 \leq k \leq L$ in
the lattice is occupied or empty. At a given time $t$ the state of the system
is completely described by the probability distribution $f(\underline{n};t)$.
In this class of models there are ten possible reactions in addition to
right and left hopping (diffusion), so altogether one has to specify 12
independent rates $a_{ij}\geq 0$ (Tab.~1).

\begin{table}[t]
\caption{Bulk reaction and diffusion rates for nearest neighbour exclusion
processes of identical particles. The numbers $a_{ij}$ are the rate of change
of the occupation numbers $\{n_k,n_{k+1}\}$.}
\vspace{0.4cm}
\begin{center}
\begin{tabular}{|c|c|l||c|c|l|}
\hline
& & & & & \\
Process & Rate & & Process & Rate & \\ \hline
& & & & & \\
01 $\rightarrow$ 10 & $a_{32}$ & diffusion &
10 $\rightarrow$ 01 & $a_{23}$ & diffusion \\
\hline
& & & & & \\
11 $\rightarrow$ 01 & $a_{24}$ & coagulation &
01 $\rightarrow$ 11 & $a_{42}$ & decoagulation \\
11 $\rightarrow$ 10 & $a_{34}$ & coagulation &
10 $\rightarrow$ 11 & $a_{43}$ & decoagulation \\
\hline
& & & & & \\
00 $\rightarrow$ 01 & $a_{21}$ & creation &
01 $\rightarrow$ 00 & $a_{12}$ & annihilation \\
00 $\rightarrow$ 10 & $a_{31}$ & creation &
10 $\rightarrow$ 00 & $a_{13}$ & annihilation \\
00 $\rightarrow$ 11 & $a_{41}$ & pair creation &
11 $\rightarrow$ 00 & $a_{14}$ & pair annihilation \\
& & & & & \\ \hline
\end{tabular}
\end{center}
\end{table}

The stochastic dynamics are defined by the master equation
\be
\frac{d}{dt} f(\underline{n};t) = \sum_{\underline{n}'}
\left[ w(\underline{n};\underline{n}') f(\underline{n}';t) -
w(\underline{n}';\underline{n}) f(\underline{n};t) \right]
\ee
where the reaction-diffusion rates $w(\underline{n};\underline{n}')$
for a change from configuration $\underline{n}' \rightarrow \underline{n}$
are equal to the sum\\
$\begin{array}{rcl}
\displaystyle \sum_{k=1}^L \left\{ \delta_{n_k',0}\delta_{n_{k+1}',0} \left[
a_{21}\delta_{n_k,0}\delta_{n_{k+1},1} + a_{31}\delta_{n_k,1}\delta_{n_{k+1},0}
+ a_{41} \delta_{n_k,1}\delta_{n_{k+1},1} \right] + \right. & & \\
\delta_{n_k',0}\delta_{n_{k+1}',1}\left[
a_{12}\delta_{n_k,0}\delta_{n_{k+1},0} + a_{32}\delta_{n_k,1}\delta_{n_{k+1},0}
+ a_{42} \delta_{n_k,1}\delta_{n_{k+1},1} \right] + & & \\
\delta_{n_k',1}\delta_{n_{k+1}',0}\left[
a_{13}\delta_{n_k,0}\delta_{n_{k+1},0} + a_{23}\delta_{n_k,0}\delta_{n_{k+1},1}
+ a_{43} \delta_{n_k,1}\delta_{n_{k+1},1} \right] + & & \\
\left. \delta_{n_k',1}\delta_{n_{k+1}',1}\left[
a_{14}\delta_{n_k,0}\delta_{n_{k+1},0} + a_{24}\delta_{n_k,0}\delta_{n_{k+1},1}
+ a_{34} \delta_{n_k,1}\delta_{n_{k+1},0} \right] \right\} . & & \\
 & &
\end{array}$

This somewhat lengthy expression becomes more compact in the quantum
Hamiltonian formalism (\ref{1}): To each configuration
$\underline{n}$ a vector $|\,\underline{n}\,\rangle$ which, together with the
transposed vectors $\langle\,\underline{n}\,|$, form an orthonormal basis
of $({\bf C}^2)^{\otimes L}$. In spin language this corresponds to a mapping to
a spin 1/2 chain by identifying a vacancy (particle) at site $k$ with spin up
(down) at this site. The probability distribution is then given by the
state vector $|\,f(t)\,\rangle=\sum_{\underline{n}} f(\underline{n};t)
|\,\underline{n}\,\rangle$ and the formal solution of the master equation
(\ref{1}) in terms of the initial distribution $|\,f(0)\,\rangle$ is given
by $|\,f(t)\,\rangle = \exp(-Ht) |\,f(0)\,\rangle$.
The stochastic dynamics are defined by the master equation (\ref{1})
with\cite{Schu95}
\be
H = \sum_{k=1}^L h_k
\label{2}
\ee
where the matrices $h_k$ act non-trivially only on sites $k,k+1$ and are
given by
\be h_k = - \left(
\begin{array}{cccc}
a_{11} & a_{12} & a_{13} & a_{14} \\
a_{21} & a_{22} & a_{23} & a_{24} \\
a_{31} & a_{32} & a_{33} & a_{34} \\
a_{41} & a_{42} & a_{43} & a_{44}
\end{array} \right)_{k,k+1}
\label{3}
\ee
with $a_{jj}=-\sum_{\stackrel{i=1}{i\neq j}}^{4} a_{ij}$.

The connection of $H$ to the Heisenberg quantum chain becomes apparent by the
similarity transformation $\tilde{H} = \Phi VHV^{-1}\Phi^{-1}$
with $V=\exp(S^+)$ where $S^+=\sum_{k=1}^L s^+_k$ and $s^{\pm}_k=(\sigma^x_k
\pm i\sigma^y_k)/2$ are the spin lowering and raising operators acting on site
$k$ and with $\Phi=\exp({\cal E}\sum_k k \sigma^z_k)$ where ${\cal E}$ is a
suitably chosen constant \cite{Schu95}. On the ten parameter submanifold
defined by
\bea
a_{34} & = & a_{21} + a_{41} + a_{12} + a_{32}
             - a_{23} - a_{43} - a_{14} \label{4} \\
a_{24} & = & a_{31} + a_{41} + a_{13} + a_{23}
             - a_{32} - a_{42} - a_{14}.
\label{5}
\eea
the transition matrices have now the structure $\tilde{h_k} =
h^{XXZ}_k + h^-_k$. Here $h^{XXZ}_k$ commutes with $S^z = \sum_{k=1}^L
\sigma^z_k/2$ and $h^-_k$ is a sum of two parts which lower the $z$-component
of the spin on sites $k,k+1$ by one and two units respectively. So one finds
\be
\tilde{H} = H^{XXZ} + H^-
\label{6}
\ee
where $H^{XXZ}$ is the Hamiltonian of the
anisotropic Heisenberg ferromagnet with twisted boundary conditions in a
magnetic field. The crucial observation is that $H^-$ {\em does not change
the spectrum of} $H^{XXZ}$, since $H^{XXZ}$ may be block-diagonalized
into blocks with fixed quantum number $S^z$ and $H^-$ connects only blocks
of given $S^z$ with blocks with quantum numbers $S^z-1$ and
$S^z-2$.\footnote{This mechanism was first noticed in a similar context in
Alcaraz et al.\protect\cite{Alca94}.}

Quantities of interest are expectation values (i.e. $r$-point correlation
functions)
$\langle\,n_{k_1}(t) \dots n_{k_r}(t)\,\rangle_{f_0} =
\langle\,s\,|n_{k_1} \dots n_{k_r} e^{-Ht} |\,f(0)\, \rangle$ which give the
probability of finding particles on the set of sites $\{k_1, \dots , k_r\}$
at time $t$, if the initial distribution at time $t=0$ was $f_0$.
Here $\langle \; s\,| = \sum_{\underline{n}} \langle\,\underline{n}\,|$
and $n_k = (1-\sigma^z_k)/2$ is the projector on states with a particle on site
$k$.
>From the Bethe ansatz one finds now that the spectrum has an energy gap (i.e.
inverse
correlation time) $\mu' = 4 a_{41} + 2(a_{21}+a_{31}) +a_{12}+a_{13} - a_{42} -
a_{43}
\geq 0$. If $\mu' = 0$ the dynamical exponent turns out to be $z=2$.
Note also that $V$ transforms a $r$-point density correlation function
into a matrix element in the sector with $r$ down spins. Since
$H^-$ only creates down spins, only transformed initial states with
$l \leq r$ down spins will contribute to the correlation function.
This surprising
simplification allows for an exact calculation of the local average density
for any initial state even though we are dealing with a non-trivial
interacting many particle system \cite{Schu95}.

\section{The Dynamic Matrix Ansatz}

The results of the last section involve the constraints (\ref{4}),
(\ref{5}) and do not apply e.g. for the asymmetric exclusion process.
Also this model is integrable, but a calculation of time-dependent correlation
functions has not yet been achieved. In order to solve this problem we now
formulate a dynamic matrix ansatz for the general reaction-diffusion system
defined by (\ref{2}) and (\ref{3}), generalizing earlier work
\cite{Stin95a,Stin95b} for diffusion only. Instead of periodic
boundary conditions we consider a system with open boundaries where
particles are injected (absorbed) at site 1 with rate $\alpha$ ($\gamma$)
and at site $L$ with rate $\delta$ ($\beta$). Therefore $H=b_1+b_L +
\sum_{k=1}^{L-1}h_k$ with suitably chosen
boundary matrices $b_1,b_L$ \cite{Stin95a}.

The ansatz is to take $|\,f(t)\,\rangle = \langle\langle\,W\,|\{ \prod_{k=1}^L
(E(t) + D(t) \sigma^-_k) \} |\,0\,\rangle|\,V\,\rangle\rangle/Z_L$
where $|\,0\,\rangle$ is the state with all spins up and $D,E$ are
time-dependent matrices satisfying an algebra obtained from the master equation
(\ref{1}). The (time-independent) vectors $\langle\langle\,W\,|$ and $|\,V\;
\rangle\rangle$ on which $D$ and $E$ act are determined from the boundary
terms in the master equation and
$Z_L=\langle\langle\,W\,|C^L|\,V\,\rangle\rangle$ where $C=D+E$ is a
normalization.
In this framework the $r$-point density correlation function is given by
$\langle\,n_{k_1}(t) \dots n_{k_r}(t)\,\rangle_{f_0} =
\langle\langle\,W\,|C^{k_1-1}DC^{k_2-k_1-1}D\dots
C^{L-k_r}|\,V\,\rangle\rangle/Z_L$.
Therefore, given a matrix representation of the algebra satisfied by
$D,E$, the computation of time-dependent correlation functions is reduced
to the much simpler calculation of matrix elements of a product of $L$
matrices.

It is easy to see that (\ref{1}) is solved if for each pair of sites one
satisfies
\bea
& & (\frac{1}{2} \frac{d}{dt} + h_k)(E+D\sigma^-_k)(E+D\sigma^-_{k+1})
|\,0\,\rangle \;\; =
\nonumber \\
& & \left[(S+T\sigma^-_k)(E+D\sigma^-_{k+1}) -
(E+D\sigma^-_k)(S+T\sigma^-_{k+1})\right] |\,0\,\rangle
\label{9}
\eea
where $S,T$ are auxiliary operators satisfying
\bea
\langle\langle\,W\,| \left[(\frac{1}{2} \frac{d}{dt} + b_1)(E+D\sigma^-_1)
+(S+T\sigma^-_1) |\,0\,\rangle \right] & = & 0 \label{10a} \\
\left[ (\frac{1}{2} \frac{d}{dt} + b_L)(E+D\sigma^-_L) -(S+T\sigma^-_L)
|\,0\,\rangle \right] |\,V\,\rangle\rangle & = & 0.
\label{10b}
\eea
By comparing each of the four terms in (\ref{9}) proportional to
$|\,0\,\rangle$, $\sigma^-_k|\,0\,\rangle$, $\sigma^-_{k+1}|\,0\,\rangle$
and $\sigma^-_k\sigma^-_{k+1}|\,0\,\rangle$ resp. one obtains four
quadratic relations for the operators
$D,E,S,T$. Eqs. (\ref{10a}) and (\ref{10b})
give two pairs of equations which define $\langle\langle\,W\,|$
and $|\,V\,\rangle\rangle$. Introducing
\bea
A^{(1)}_j & = & - (a_{21}+a_{31}+a_{41}) E^2 + a_{12} ED
                 + a_{13} DE + a_{14} D^2  \\
B^{(1)}_j & = & a_{21} E^2 - (a_{12}+a_{32}+a_{42}) ED
                 + a_{23} DE + a_{24} D^2 \\
B^{(2)}_j & = & a_{31} E^2 + a_{32} ED
             - (a_{13}+a_{23}+a_{43}) DE + a_{34} D^2 \\
A^{(2)}_j & = & a_{41} E^2 + a_{42} ED
             + a_{43} DE - (a_{14}+a_{24}+a_{34}) D^2.
\eea
one finds
\bea
\label{10c}
\frac{1}{2}\frac{d}{dt} E^2 - [S,E]  & = & A^{(1)} \\
\label{10d}
\frac{1}{2}\frac{d}{dt} ED - SD + ET & = & B^{(1)} \\
\label{10e}
\frac{1}{2}\frac{d}{dt} DE - TE + DS & = & B^{(2)} \\
\label{10f}
\frac{1}{2}\frac{d}{dt} D^2 - [T,D]  & = & A^{(2)}
\eea
and
\bea
\langle\langle\,W\,|
\left\{\frac{1}{2}\frac{d}{dt} E - \alpha E + \gamma D + S \right\} & = & 0 \\
\langle\langle\,W\,|
\left\{\frac{1}{2}\frac{d}{dt} D + \alpha E - \gamma D + T \right\} & = & 0 \\
\left\{\frac{1}{2}\frac{d}{dt} E - \delta E + \beta D - S \right\}
|\,V\,\rangle\rangle& = & 0 \\
\left\{\frac{1}{2}\frac{d}{dt} D + \delta E - \beta D - T \right\}
|\,V\,\rangle\rangle& = & 0 .
\eea

One may reduce this algebra by assuming that $C$ is time-independent and has
a representation where it is invertible. Eqs. (\ref{9}) then imply $[C,S+T] =
0$
and (\ref{10a}),(\ref{10b}) imply
$\langle\langle\,W\,|(S+T)=0=(S+T)|\,V\,\rangle\rangle$.
This can be solved by assuming $S+T=0$, which, as I would like to stress, is
{\em not} the most general choice. Now one can express $S$ in terms of $C$ and
$D$ and is left with only two further relations to be satisfied by $D$ and $C$
and two relations defining $\langle\langle\,W\,|$ and $|\,V\,\rangle\rangle$.
In particular, if (\ref{5}) and (\ref{6}) are satisfied, there is one relation
involving $\dot{D}$ which is linear in $D$ and one relation quadratic in $D$.
For the symmetric exclusion model $a_{23}=a_{32}=1/2$ this dynamic algebra
yields eigenvalue equations for the corresponding $XXX$-Hamiltonian with
integrable,
but non-diagonal, symmetry breaking boundary fields\cite{Stin95a,Stin95b}.
However, no
matrix representation has been found yet. This raises the question whether
non-trivial representations do exist at all.

As I will show here for the first time, the answer to this question is yes, at
least with some restrictions on the injection and absorption rates. Choosing a
basis where $C$ is diagonal one finds the representation
\bea
C \; = \; \left(
\begin{array}{cc}
 1 & 0 \\
 0 & c
\end{array} \right)
& , &
D \; = \; \left(
\begin{array}{cc}
 d & \lambda e^{-\epsilon t} \\
 0 &  cd
\end{array} \right).
\label{15}
\eea
with $\epsilon=(\alpha+\beta+\gamma+\delta)/2$,
$c=1-\alpha-\gamma=(1-\beta-\delta)^{-1}$, $d=\alpha/(\alpha+\gamma)=
\delta/(\beta+\delta)$ and $\langle\langle\,W\,|$, $|\,V\,\rangle\rangle$
arbitrary
but $\langle\langle\,W\,|\,V\,\rangle\rangle\neq 0$.
In this representation $\lambda$ is an arbitrary parameter specifying
the initial distribution. One may also use it for the construction of (right)
eigenstates of $H$, since the expression $\langle\langle\,W\,|E^{k_1-1}
DE^{k_2-k_1-1}\dots E^{L-k_r}|\,V\,\rangle\rangle$ is a superposition
of wave functions $\Psi_{\epsilon_i}(k_1,\dots,k_r)$ of eigenstates with
eigenvalues
$\epsilon_i$. The argument is the position of $r$ down spins on
sites $k_1,\dots,k_r$. Taking $\lambda=0$ corresponds to taking the stationary
distribution as initial state. This is an eigenstate with energy 0. The
terms proportional to $\lambda$ give the wave function for an eigenstate
with energy $\epsilon$. The quantity $1/(\ln{|c|})$ plays the role of a
spatial correlation length.

\section{Non-equilibrium spin relaxation}

A phenomenon of wide interest in physics and chemistry is the growth
of domains in non-equilibrium two-phase systems. The best-known example
is perhaps the Ising model with domains of up- and down spins, separated by
domain walls. The energy of the Ising model is given by the nearest neighbour
sum $E = - J \sum s_i s_j$. Since the creation of a local domain wall costs
an energy $J$ the system tries to organize itself at low temperature
into large domains
of uniform magnetization. Starting from a high-temperature equilibrium
state with many domain walls and quenching to low temperatures
leads to a coarsening process: Small domains
of uniform magnetization merge to form larger domains since then the total
length of the domain walls and thus the energy decreases.

Glauber \cite{Glau63} introduced spin-flip dynamics which ensure that
the system reaches the equilibrium distribution at temperature $T = 1/\beta$
of the one-dimensional
Ising model. In this model a spin within a domain of
equal magnetization is flipped with a rate $\mu = 1 - \tanh{\beta J}$,
whereas a spin in a region of opposite magnetization is flipped
with a rate $\lambda = 1 + \tanh{\beta J}$. At domain boundaries spins
are flipped with unit rate, since no change in energy is involved. This
process can be visualized in the following way:
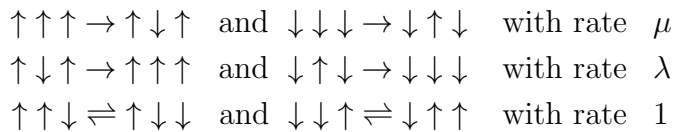
\begin{eqnarray}
\uparrow \; \uparrow \; \uparrow \; \to \; \uparrow \; \downarrow \; \uparrow
\;
\mbox{ and }
\downarrow \; \downarrow \; \downarrow \; \to \; \downarrow \;  \uparrow \;
\downarrow \; & \mbox{with rate} & \mu \nonumber \\
\uparrow \; \downarrow \; \uparrow \; \to \; \uparrow \; \uparrow \; \uparrow
\;
\mbox{ and }
\downarrow \; \uparrow \; \downarrow \; \to
 \; \downarrow \;  \downarrow \;
\downarrow \; & \mbox{with rate} & \lambda \nonumber \\
\uparrow \; \uparrow \; \downarrow \;
\rightleftharpoons \;
\uparrow \; \downarrow \; \downarrow \;
\mbox{ and }
\downarrow \; \downarrow \; \uparrow \;
\rightleftharpoons \; \downarrow \;  \uparrow \;
\uparrow \; & \mbox{with rate} & 1 \nonumber
\end{eqnarray}

Glauber dynamics can also be seen as a reaction-diffusion system. One simply
identifies an up-spin with a vacancy and a down-spin with a particle. In
one dimension at zero temperature the process can then be described as follows:
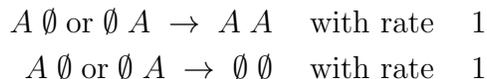
\begin{eqnarray}
A \; \emptyset \; \mbox{or} \; \emptyset \; A \; \to  \; A \; A
& \mbox{ with rate } & 1 \nonumber \\
A \; \emptyset \; \mbox{or} \; \emptyset \; A \; \to \;\emptyset\;\emptyset
& \mbox{ with rate } & 1 \nonumber
\end{eqnarray}
This can obtained by a translational rearrangement of the three-site
interactions in terms of two-site processes. One realizes then that Glauber
dynamics can be represented by a stochastic Hamiltonian of the form (\ref{2}),
(\ref{3}) with $a_{12}=a_{13}=a_{42}=a_{43}=1$. Furthermore, these rates
satisfy the constraints (\ref{5}). We stress that this relation to a
reaction-diffusion system is not really a mapping, but just a
certain choice of language which we use in order to make contact
with Sections 2 and 3.
There are two different non-trivial mappings \protect\cite{Racz85,Henk95} to
the
process of diffusion-limited annihilation which has little in common with
the process described here. These mappings are useful as they show that
Glauber dynamics can be described and solved in terms of free fermions
\cite{Sigg77,Alca94,Lush87,Gryn94,Schu95b,Sant97a} and that
the associated quantum chain is a non-hermitian anisotropic
Heisenberg chain related to the seven-vertex model.

The Glauber relaxation rules involve single spin flips and thus do not
conserve the total magnetization. Kawasaki \cite{Kawa66} introduced
spin-exchange dynamics which also lead to an equilibrium Ising
distribution, but which {\em do} conserve the total magnetization. At
infinite temperature, these dynamics reduce to simple exchange of
neighbouring spins with some rate $a_{23}=a_{32}=\zeta$, i.e. to the
symmetric exclusion process described above. Due to the lack general
theorems on the dynamics of non-equilibrium systems it is now of interest
to investigate a spin system the behaviour of which results from a coupling
to two heat baths at different temperatures - one leading to zero-temperature
Glauber dynamics, the other to infinite-temperature Kawasaki dynamics
\cite{Droz89}. In such a situation there is a competition: The diffusion
process tries to disorder the system, while the Glauber process tries to create
an ordered system of uniform structure. Hence the questions arise, which
process wins, and how is the stationary state reached.

In one dimension this problem was addressed by studying the dynamical
spin-spin (= particle-particle) correlations for a translationally
invariant initial state \cite{Droz89}, using the fact that the equations of
motion for correlation functions decouple into closed subsets. By solving
the equations Droz et al. could show that at any (finite) value $\zeta$ of
the coupling strength the system orders and that the spin-spin correlation
function
behaves at large times like the zero-temperature Glauber correlator.
Here we use the matrix product ansatz to prove non-perturbatively that
this remains true for {\em all} correlation functions of finite order.
The leading contribution to time-dependent correlation functions
(for large times) is always the zero-temperature Glauber correlation
function \cite{Feld70}. Corrections resulting from the coupling to the
infinite-temperature heat bath are of subleading order $1/\sqrt{t}$
(relative to the leading contribution).\footnote{We mention in
passing that this process is a simple toy model of growing tissue
cell populations \cite{Dras96}. The decoagulation process
$A \emptyset,\emptyset A \to AA$ with unit rate describes cell division, while
the particle hopping with rate $\zeta$ corresponds to the diffusive motion of
cells in their environment. In addition to that we allow for a death process
$A \emptyset,\emptyset A \to \emptyset\emptyset$
with rate $q$ which kills both the original cell and its offspring during the
decoagulation (cell mitosis). It is intuitively clear that for $q<1$ the cell
population will grow until all space is covered, while for $q>1$ the population
will eventually die out. Therefore it is of interest to study the case
$q=1$ when creation of offsprings and the death process balance each other.
The process leads to an ordered state also in three dimensions \cite{Schu98b}.
This implies that either all tumor cells die, or, with equal probability, cover
the whole available space.}

To prove our assertion we consider the matrix algebra describing the process.
Since we are more interested in spin
variables we introduce $\tilde{D} = C - 2D = E-D$.
We restrict ourselves to translationally invariant initial states.
Because of translational invariance time-dependent spin
expectation values are given by a trace over matrices
\be
\langle\,\sigma^z_{k_1}(t) \dots \sigma^z_{k_r}(t)\,\rangle_{f_0} =
\mbox{Tr }\{ C^{k_1-1}\tilde{D}C^{k_2-k_1-1}\tilde{D}\dots C^{L-k_r}\}/Z_L
\ee
where $Z_L = \mbox{Tr }C^L$.
We reduce the algebra (\ref{10c}) - (\ref{10f}) as in the case of the
symmetric exclusion process by setting $S+T=0$. Eliminating $S$
leads then to the algebra generated by $C,\tilde{D}$ and $C^{-1}$ with the
relations $CC^{-1}=C^{-1}C=1$, $d/(dt) C = 0$ and
\bea
\label{16}
\frac{d}{dt}\tilde{D} & = & (1+\zeta) \left( C\tilde{D}C^{-1} +
C^{-1}\tilde{D}C - 2\tilde{D} \right) \\
\label{17}
2(1-\Delta)  & = & \tilde{D}C^{-1}\tilde{D}C^{-1} +
C^{-1}\tilde{D}C^{-1}\tilde{D} - 2 \Delta C^{-1}\tilde{D}^2 C^{-1}
\eea
with the constant $\Delta = \zeta/(1+\zeta)$ determined by coupling ratio
$\zeta$. The $r$-point spin correlation function can now be rewritten
\bel{18}
\exval{\sigma^z_{k_1}(t)\dots \sigma^z_{k_r}(t)} =
\mbox{Tr }\{ \tilde{D}_{k_1} \dots \tilde{D}_{k_r} C^L\} /Z_L
\ee
where $\tilde{D}_k=C^{k-1} \tilde{D} C^{-k}$ and
$k_{i+1} > k_i$. These relations provide an alternative, purely
algebraic definition of the spin relaxation process.

Relation (\ref{16}) is linear in $\tilde{D}$ and we procede by constructing the
{}Fourier transforms ${\cal D}_p = \sum_k e^{ipk} \tilde{D}_k$ to reformulate
the algebra in terms of the Fourier components. Since
\begin{equation}\label{19}
C {\cal D}_p C^{-1} = e^{-ip} {\cal D}_p ,
\end{equation}
the time-dependence of ${\cal D}_p$ is now simply obtained from (\ref{16})
\bel{20}
{\cal D}_p(t) = e^{-\epsilon_p t} {\cal D}_p(0)
\ee
in terms of the initial matrix ${\cal D}_p(0)$ and the "energy"
\bel{21}
\epsilon_p = 2(1+\zeta) (1-\cos{p}).
\ee

>From (\ref{17}) follows
$2(1-\Delta)\delta(p) = \int dp' {\cal D}_{p'} {\cal D}_{p-p'}
(1+e^{-ip}-2\Delta e^{ip'-ip})$. Since this relation holds for all times
(and all $p$), the integral can be divided
into separate time-components, each of which must vanish.
If $p_1$ and $p_2$ are non-zero this leads to
\bel{22}
{\cal D}_{p_1}{\cal D}_{p_2} = S(p_2,p_1) {\cal D}_{p_2}{\cal D}_{p_1}
\ee
with the two-body scattering matrix
\begin{equation}\label{23}
 S(p_2,p_1) = - \frac{1+e^{i p_1 + ip_2}-2 \Delta e^{i p_2}}
{1+e^{i p_1 + ip_2}-2 \Delta e^{i p_1}}
\end{equation}
known from the usual anisotropic Heisenberg chain \cite{Yang66}.

This relation for ${\cal D}_p$ is derived for the dynamic
components with $p \neq 0$. Hence the static term in the l.h.s. of
(\ref{17}) does not reappear in (\ref{22}), but in a different relation
involving the static parts ${\cal D}_0=\sum_n C^{n-1} D C^{-n}$ and
${\cal I} = \sum_n n C^{n-1} D C^{-n}$. These quantities need separate
treatment in a similar way as the static components of the operators for
the symmetric exclusion process with open boundaries \cite{Schu98,Sant97b}.
However, since the stationary state (all spins up or all spins down) is
not interesting we do not consider the static Fourier components.

The momentum space formulation of the algebra provides another equivalent
formulation of the process. To calculate expectation values we do not search
for a representation of this algebra but use a different strategy.
In terms of the Fourier components the correlator (\ref{18}) reads
\bel{24}
\exval{\sigma^z_{k_1}(t)\dots \sigma^z_{k_r}(t)} =
\left( \prod_{i=1}^m \int \frac{d p_i}{2\pi}
e^{-p_ik_i-\epsilon_{p_i}t} \right)
T(\{p_i\})
\ee
where the so far undetermined matrix element
\bel{25}
T(\{p_i\})= \mbox{Tr }\{{\cal D}_{p_1}(0) \dots{\cal D}_{p_m}(0)C^L\} / Z_L
\ee
depends only on the initial distribution.

{}First consider the one-point function $\exval{\sigma^z_k(t)}$. Because of
translational invariance, this local magnetization is independent of
space. This is reflected in the invariance of the trace under cyclic
permutation. Thus (\ref{25}) for the one-point function together with
(\ref{19}) imply $e^{ip} = e^{2ip} = e^{3ip} = \dots = e^{ipL} =1$ and
therefore $T(p) = 2 \pi c \delta(p)$. This yields $\exval{\sigma^z_k(t)}=c$
for all times and restates nothing but the known result that the average
magnetization remains constant under the time evolution of the spin relaxation
model \cite{Droz89}. The initial magnetization $m_0$ fixes $c=m_0$.

We obtain a non-trivial result for the two-point correlator. We use (\ref{19}),
(\ref{23}) and the cyclic property of the trace and find
\bea
T(p_1,p_2) & = & \mbox{Tr }\{{\cal D}_{p_1}(0) {\cal D}_{p_2}(0)C^L\} / Z_L
\nonumber \\
& = & S(p_2,p_1) \mbox{Tr }\{{\cal D}_{p_2}{\cal D}_{p_1} C^L \} / Z_L
\nonumber\\
& = & e^{ip_1L} S(p_2,p_1) \mbox{Tr }\{{\cal D}_{p_2} C^L {\cal D}_{p_1}\} /
Z_L
      \nonumber\\
\label{26}
& = & e^{ip_1L} S(p_2,p_1) T(p_1,p_2).
\eea
A similar cycling procedure gives a second relation \be
T(p_1,p_2) = e^{ip_2L} S(p_1,p_2) T(p_1,p_2).
\ee
Using $S(p_1,p_2) = S^{-1}(p_2,p_1)$ we conclude that $T(p_1,p_2)$ is non-zero
only if the momenta $p_{1,2}$ satisfy the Bethe-ansatz equations \cite{Yang66}
\bea
\label{27a}
e^{ip_1L} & = & S(p_1,p_2) \\
\label{27b}
e^{ip_2L} & = & S(p_2,p_1).
\eea
Translational invariance requires also $p_1+p_2 = 0$.

Moreover, the first of the equations (\ref{26}) yields a functional equation
for the matrix element
\bel{28}
T(p_1,p_2) = S(p_2,p_1) T(p_2,p_1).
\ee
This functional equation for $T(p_1,p_2)$ is solved by the Bethe wave functions
\cite{Beth31,Thac81,Gaud83}
\begin{equation}\label{29}
\Psi_{p_1,p_2}(l_1,l_2) = A(l_1,l_2)\left( e^{i p_1 l_1 + ip_2 l_2} +
S(p_2,p_1)
e^{i p_2 l_1 + ip_1 l_2}\right),
\end{equation}
for $k_2 > k_1$ with some amplitude $A(l_1,l_2)$ determined by the initial
value
of the correlator. It can be calculated from the integral representation
(\ref{24}) of the full time-dependent correlator by setting $t=0$.

{}For a
finite system the integral has to be replaced by a sum over the solutions
of the Bethe ansatz equations (\ref{27a}), (\ref{27b}). However, in an
infinite system the set of solutions becomes dense. The only subtlety
arises then from the bound states defined by the pole of the scattering
amplitude $S$. This pole corresponds to
the two-particle bound state already known from the original solution of
Bethe \cite{Beth31} for $\Delta = 1$.
One can fix the contour of integration by setting
$A(l_1,l_2) =  \exval{n_{l_1}(0) n_{l_2}(0)}$. This gives
\be
T(p_1,p_2) = \sum_{l_1,l_2} \exval{n_{l_1}(0) n_{l_2}(0)}
\left( e^{i p_1 l_1 + ip_2 l_2} + S(p_2,p_1)
e^{i p_2 l_1 + ip_1 l_2}\right)
\ee
with the sum being restricted to the domain $l_2 > l_1$.
We prescribe the appropriate contour of integration by isolating in $S$
the constant part $S_0=- 1$ which corresponds to
non-interacting fermions. One writes
\bel{30}
S(p_2,p_1) = - 1 + 2 \Delta (1-e^{ip_1-ip_2})
\int_0^\infty du \; e^{-u(e^{ip_1}+e^{-ip_2}-2 \Delta e^{ip_1-ip_2})}.
\ee
and integrates both $p_1$ and $p_2$ from $0$ to $2\pi$ along the real axis
{\em before} integrating over $u$. Both this definition of the integration
and the choice for the amplitude $A(l_1,l_2)$ ensure that the initial
condition is indeed satisfied in the physical domain
$k_2 > k_1,\;l_2 > l_1$. With the constraint $p_1+p_2=0$ originating in
translational invarinace one recovers in this way the expression for the
two-point correlator derived by Droz et al. \cite{Droz89} by direct
solution of the equations of motion with generating function techniques.

Note that for the pure Glauber case $\Delta = 0$ and hence $S = S_0 = -1$.
In this case the ${\cal D}_p$ anticommute and can represented e.g. by
a Jordan-Wigner transformation of the usual local spin-1/2 raising or lowering
operators $s^\pm_k$. We recover in a surprising way the free fermion nature of
this process. However, since Glauber dynamics are well-understood \cite{Feld70}
we do not further pursue this matrix representation.
In the present context we are more interested in the observation that for
large times only small $p_1,p_2$ contribute to the integral (\ref{24}).
Making a substitution of variables $p_i \to p_i/\sqrt{t}$ and expanding
for large $t$ leads to
\bel{31}
S = - 1 + O(t^{-1/2})
\ee
for the late time behaviour of the scattering amplitude $S$ (\ref{30}).
This proves our assertion for the two-point function: For large times
the correction to the correlator due to coupling to the infinite-temperature
Kawasaki heat bath is of subleading order $O(t^{-1/2})$ for any finite
coupling $\zeta$.

Higher order correlation functions are treated analogously. The permutation
of ${\cal D}_p$ matrices using the matrix relations (\ref{19}), (\ref{23})
yields the Bethe ansatz equations and functional equations for the matrix
elements
$T(\{p_i\})$. For a $r$-point correlator this functional equation is solved
by the $r$-particle Bethe wave function \cite{Beth31,Yang66}. Because of the
integrability the scattering amplitudes factorize into products of two-body
amplitudes. Thus for large times the leading part comes from the free-fermion
amplitude $S=-1$ and therefore the leading part of the correlator is
independent of the Kawasaki coupling $\zeta$ and given by the pure Glauber
correlation function. The leading correction which results
from this coupling is of order $t^{-1/2}$. This proves our assertion.

\section{Conclusions}

A 10-parameter class of stochastic reaction-diffusion systems can be mapped
by a similarity transformation to a generalized Heisenberg
quantum chain, the spectrum of which can be obtained by the Bethe ansatz.
It turns out that time-dependent $r$-point density correlation functions
are given by the $l\leq r$-magnon sectors. As an alternative to that
approach we introduced a dynamic matrix ansatz for the general 12-parameter
model. This ansatz reduces the calculation of all correlators to the
calculation
of certain matrix elements. These matrices satisfy an infinite-dimensional
algebra which is determined by the bulk dynamics of the process. The
boundary conditions determine which matrix elements one has to take.

In the cases of the symmetric exclusion process and of the non-equilibrium
spin relaxation model of Droz et al. the algebra satisfied by the matrices
can be used to obtain the spectrum of the corresponding quantum
Hamiltonian. We constructed a two-dimensional time-dependent matrix
representation of the algebra for the symmetric exclusion process from which
one obtains explicitly all $r$-point density correlators for a
one-parameter class of initial states. The corresponding eigenvectors of the
Heisenberg chain are the ground state with energy 0 and a bound state with
energy $\epsilon=(\alpha+\beta+\gamma+\delta)/2$. An alternative treatment
of the dynamical algebra exploits directly the algebraic
relations which result in functional equations for the dynamical
part of the correlator. We have solved these equations for the spin
relaxation model in terms of Bethe wave functions.
We proved that independently of the coupling to the infinite-temperature
heat bath all $r$-point equal-time correlation functions decay to leading
order in time like the zero-temperature Glauber correlators.

>From a mathematical point of view the dynamical matrix algebra and its
representation theory is not yet well-understood. The stationary version
of the extended algebra was considered by Hinrichsen et al. \cite{Hinr96c}
who constructed a four-dimensional representation for a
coagulation/decoagulation model.
Krebs and Sandow \cite{Kreb97} could prove that the stationary algebra
extended with the auxiliary matrices forms an equivalent formulation of
the stationary master equation. This guarantees the existence of a
representation for the general case. To date there is no equivalent
theorem for the dynamical algebra and no representation theory.
A second important question concerns the relationship
between the integrability of quantum chains and the dynamic matrix ansatz
which emerged here and in other work \cite{Stin95b,Sasa97a,Sasa97b}.
The result of Krebs and Sandow for the stationary algebra shows that
there is no general relationship between the possibility of an algebraic
description and the integrability of a system. Yet for integrable models
the algebra is powerful enough not only to recover the known Bethe ansatz
equations but also to obtain results which cannot be obtained
using standard Bethe ansatz techniques. This suggests that matrix algebras
describing integrable models have special, as yet undiscovered
properties.

\section*{Acknowledgments}
I would like to thank D. Drasdo,
V. Hakim, M. Evans, Z. R\'acz and R. Stinchcombe for useful and stimulating
discussions.

\end{document}